# A Theory-driven and AI-enhanced Simulation Platform for Cultivating Nutrition Literacy

**Shan Li**[*]
College of Health / College of Education
Lehigh University
Bethlehem, PA, USA, 18015
Email: shla22@lehigh.edu
ORCID: https://orcid/org/0000-0001-6001-1586

**Guozhu Ding**
Department of Educational Technology
Guangzhou University
Guangzhou, China, 510165
Email: dinggz@gzhu.edu.cn
ORCID: https://orcid.org/0000-0002-2043-8320

**[*]Corresponding Author**
All correspondence should be sent to: Shan Li, HST Building 132, 124 E. Morton Street, Lehigh University, Bethlehem, PA 18105, USA. Email: shla22@lehigh.edu.

## Declarations

*Disclosure Statement*
The authors report that there are no competing interests to declare.

*Ethics Approval Statement*
This study received an exemption from the Institutional Review Board (IRB) at Lehigh University, where the research was conducted.

*Acknowledgements*
This work is supported by Lehigh University internal grant under grant number FRGS00011434. Any opinions, findings, and conclusions or recommendations expressed in this paper, however, are those of the authors and do not necessarily reflect the views of Lehigh University.

## Abstract

**Objective**: This study introduced a theory-driven, AI-enhanced simulation platform for enhancing nutrition literacy. We tested its effectiveness in supporting learners as they made decisions about food and beverage choices through authentic scenario-based learning experiences.

**Methods**: A mixed-methods study was conducted with 114 university students who completed two learning scenarios using the platform. Data collection included quantitative ratings of usefulness and ease of use, along with qualitative feedback through open-ended responses. Thematic analysis was employed to identify key patterns in user experiences.

**Results**: The platform received high ratings for both usefulness and ease of use. Qualitative analysis revealed four main strengths: interactive learning approach, scenario authenticity, AI assistance for understanding nutritional information, and practical comparison tools for decision-making.

**Conclusions and Implications**: This research advances understanding of AI-enhanced learning environments for researchers, provides practical guidance for health educators implementing technology-based interventions, and informs policymakers about effective approaches to improving population-level nutrition literacy. (150)

**Introduction**

The profound impact of health literacy on individual well-being and societal outcomes is well-documented. Individuals with limited health literacy face numerous challenges, including increased hospitalizations, poorer management of chronic diseases, and decreased likelihood of receiving preventive care.[1] Notably, nearly half of American adults struggle with comprehending and acting upon health information,[2,3] highlighting a critical need for innovative interventions to enhance health literacy across diverse populations.

One domain where health literacy deficits are particularly concerning is nutrition. Individuals with limited nutrition literacy tend to check food labels less frequently and demonstrate difficulties in understanding nutrition information.[4,5] Consequently, they are more likely to make suboptimal dietary choices, contributing to the alarming prevalence of obesity and diet-related chronic diseases.[6] Alarmingly, these issues manifest from an early age, underscoring the importance of cultivating nutrition literacy during childhood and adolescence to establish lifelong healthy habits.[7]

Traditional approaches to health and nutrition education often rely on passive information delivery, which fails to actively engage learners in developing practical decision-making skills.[8,9] While research has demonstrated the effectiveness of experiential learning in various educational contexts, few studies have examined how technology-enhanced simulations can support the development of nutrition literacy.[10,11] This gap is particularly notable given the increasing role of technology in education and healthcare decision-making.

To address these challenges, we introduce Healthy Choice, an innovative computer simulation platform that helps cultivate individuals' nutrition literacy. The platform creates an immersive learning environment where users analyze realistic nutritional scenarios, evaluate product information, and make evidence-based recommendations. This approach aims to develop both nutrition knowledge and essential decision-making skills through structured, interactive experiences. This study also examines the effectiveness of the Healthy Choice platform through a mixed-methods evaluation involving university students. By understanding how learners interact with and benefit from the platform, we aim to advance the design of technology-enhanced nutrition education interventions.

This research contributes to both theoretical understanding and practical implementation of nutrition education. The findings have implications for researchers studying technology-enhanced learning environments, educators developing health literacy interventions, and policymakers working to improve population-level nutrition literacy. Through systematic evaluation of the Healthy Choice platform, we provide insights into how interactive, AI-enhanced simulations can support the development of critical health decision-making skills.

**Methods**

This section describes our methodological approach to developing and evaluating the Healthy Choice platform. We begin by detailing the theory-driven design principles and features of the platform, followed by information about study participants and our data collection procedures. Finally, we present our analytical approach for assessing user experiences and platform effectiveness. We aim to provide a comprehensive understanding of both the platform's design rationale and its empirical evaluation.

**Theory-driven Design of the Healthy Choice Platform**

The Healthy Choice platform represents an innovative approach to nutrition education, integrating multiple learning theories to create an effective and engaging learning environment. Each feature of the platform is carefully designed based on established theoretical frameworks to support learners in developing nutrition literacy and decision-making skills.

At its core, the platform employs scenario-based learning to create authentic educational experiences. Learners assume the role of health professionals tasked with recommending suitable food and beverage products for virtual clients with specific needs and preferences. These scenarios span diverse contexts, from selecting energy drinks for study sessions to choosing sports drinks for athletic events. This approach is grounded in situated learning theory, which emphasizes that learning is most effective when embedded in authentic contexts.[12] The realistic scenarios help learners connect abstract nutritional concepts to practical applications. Additionally, the varying complexity of scenarios aligns with deliberate practice theory, which suggests that expertise develops through systematic, focused practice with progressively challenging tasks.[13]

When reviewing each scenario, learners can highlight essential information that can be captured and stored in a dedicated tracking panel. This panel serves as a metacognitive tool,[14] allowing learners to monitor their progress and reflect on the key requirements they have identified throughout their decision-making process.

Learners must critically evaluate a database of real-world product options to make informed recommendations (see Figure 1). Each product includes detailed nutritional information, ingredient lists, and marketing claims. This authentic product evaluation mirrors the challenges of sifting through abundant product information in real-life situations, requiring learners to analyze nutritional data, identify potential health impacts, and assess alignment with the virtual client's needs.

Figure 1 *A Database of Real-world Product Options*

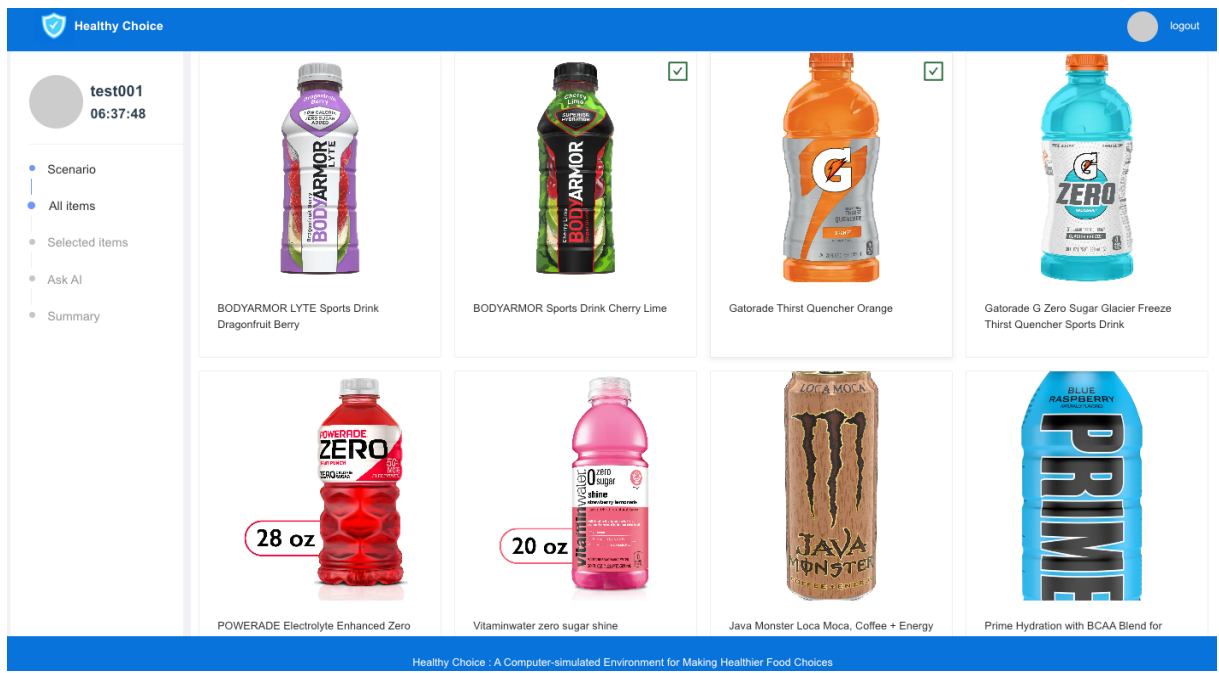

When examining each product, learners engage in two key evaluation processes that align with self-regulated learning (SRL) theory[15]: first, they assess the product's suitability by analyzing its nutritional information, ingredients, and characteristics against the scenario requirements, and second, they make an initial decision about whether to consider this product for further comparison, which prompts them to actively engage in strategic planning and decision-making. These dual evaluation steps exemplify how the platform scaffolds metacognitive monitoring and control processes, as learners must systematically gather evidence about each product's appropriateness before deciding whether it merits additional consideration in their final recommendation. Figure 2 depicts a user checking a product and making initial decisions.

Figure 2 *Checking a Specific Product for Decision-making*

[Figure: Screenshot of the Healthy Choice platform interface showing the BODYARMOR LYTE Sports Drink Dragonfruit Berry product page, with annotations: "Assess the suitability of the product for the given scenario" (pointing to the Assessment panel with options Not Appropriate, Appropriate, Highly Appropriate, Not Sure); "Click to enlarge the pictures for product details"; "Make an initial decision on whether to consider this product or not" (pointing to Initial Decision panel with Select/Not Select); and "Information support empowered by generative AI" (pointing to the question input field).]

A distinguishing feature of the platform is its integration of artificial intelligence, specifically leveraging the ChatGPI API, as a learning support tool. Learners can engage in natural conversations with this AI assistant, which draws upon its vast knowledge base to provide instant information, explain complex nutritional concepts, and guide users through their decision-making journey. The implementation of ChatGPT's capabilities exemplifies scaffolding theory in action,[11] as the AI adapts its support to each learner's unique needs and queries. When students encounter unfamiliar ingredients or need clarification about nutritional values, they can immediately consult the AI assistant, receiving personalized explanations that help bridge their knowledge gaps.

In the final stage, the platform shows learners the key information they have collected and their final recommendation, and it asks them to write a detailed justification that connects their analysis to their choice. This structured reflection step, which displays the highlighted requirements next to the selected product, helps learners explain their reasoning clearly while demonstrating how their decision aligns with the identified needs.

Together, these features create a complete self-regulated learning cycle that supports the development of nutrition literacy and decision-making skills.[15] Learners begin by analyzing scenario requirements and setting goals (forethought), proceed through systematic information gathering and decision-making (performance), and conclude with structured reflection and justification (self-reflection). The Healthy Choice platform transforms nutrition education from passive information delivery to active skill development. It could effectively support the development of critical thinking, decision-making, and health literacy skills essential for navigating today's complex nutritional landscape.

**Participants**

The evaluation study received ethical approval from the Institutional Review Board before commencement. Participants were recruited from a private university located in the northeastern United States. The study included 114 participants with an average age of 21.72 years, ranging from 17 to 35 years old. There were 74 female students (64.9%), 38 male students (33.3%), and 2 nonbinary students (1.8%). In terms of racial and ethnic background, there were 41 Asian participants (36%), 39 White/Caucasian participants (34.2%), 19 Hispanic participants (16.7%), 10 Black or African American participants (8.8%), and 1 participant identifying as Asian, White, and Pacific Islander (0.9%). The participants represented various academic disciplines, including STEM fields (Computer Science, Engineering, Mathematics), health-related programs (Population Health, Psychology, Public Health), and business-related majors (Finance, Accounting), with academic levels ranging from first-year undergraduates to graduate students and one staff member. This diverse participant pool represents a broad cross-section of the university community, with varying levels of academic experience and different disciplinary backgrounds.

**Procedures**

The study was conducted in a controlled laboratory environment with one-on-one sessions lasting approximately one hour per participant. Each session began with the presentation of a detailed consent form that outlined the study's purpose, potential benefits and risks, and emphasized the voluntary nature of participation. Participants were informed that they could withdraw from the study at any time without penalty. After providing informed consent, participants completed a pre-survey gathering demographic information.

Before engaging with the scenarios, participants received a comprehensive introduction to the computer-based simulation environment. Moreover, they were given the opportunity to practice with a sample scenario to ensure familiarity with the platform's features and functionality. This orientation phase helped participants understand how to navigate the interface, use the AI assistance feature, and complete the decision-making tasks.

Following the practice phase, participants completed two main scenarios where they were tasked with recommending appropriate food and beverage choices for virtual customers with specific needs and preferences. Each scenario required participants to analyze nutritional information, consider client requirements, and provide justified recommendations. After completing both scenarios, participants finished the session by completing a post-questionnaire to evaluate their experience with the platform.

**Data Collection**

The evaluation of the Healthy Choice platform employed a mixed-methods approach to gather comprehensive user feedback. The survey instrument was designed to capture both quantitative metrics and qualitative insights regarding the platform's effectiveness and usability. The quantitative component consisted of two key scaled questions. The first assessed participants' perceived value of the platform in developing nutrition literacy, using a scale from 1 (not valuable at all) to 10 (extremely valuable). The second measured the platform's ease of use, with participants rating from 1 (not easy at all) to 10 (extremely easy). These scaled questions provided clear metrics for evaluating the platform's effectiveness and usability. To capture deeper insights into user experiences, the survey included an open-ended question prompting participants to share their thoughts on the platform's strengths and potential areas for improvement. This qualitative component allowed participants to express their perspectives freely, which often revealed insights that might not have emerged through structured questioning alone. The survey was administered immediately after completing two learning scenarios, allowing participants to provide feedback while their experience was fresh in their minds. The resulting dataset comprised 113 quantitative ratings for each scaled question and extensive qualitative feedback from 98 participants. The collection of user responses provided a strong foundation for subsequent analysis and identification of key themes in user experience and platform effectiveness.

**Data Analysis**

The analysis approach integrated quantitative and qualitative methods to develop a comprehensive understanding of user experiences with the Healthy Choice platform. For the quantitative data, we conducted a descriptive statistical analysis of the usefulness and ease of use ratings, calculating central tendency measures (mean, median, and mode) and examining rating distributions. The qualitative data analysis employed a thematic analysis approach to identify recurring patterns and key themes in user feedback. We reviewed all 98 written responses thoroughly, coding significant statements that reflected users' experiences and perspectives. These initial codes were then grouped into broader categories based on shared meanings and relationships. Through iterative analysis and refinement, we identified several major themes: the platform's interactive approach to nutrition education, the authenticity of learning scenarios, the value of AI assistance, and the utility of comparison features. The integration of quantitative metrics with qualitative insights provided a rich, multifaceted understanding of user experiences and platform effectiveness.

**Results**

**Usefulness**

The platform's usefulness was evaluated by 113 participants on a scale from 1 to 10. The distribution of ratings shows a strong positive skew, with a mean rating of 8.19, median of 8, and mode of 8. This alignment of central tendency measures suggests a consistent and reliable assessment of the platform's value. A detailed examination of the rating distribution reveals that 24 users (21.2%) gave the platform the highest possible rating of 10, while 28 users (24.8%) rated it 9, and 31 users (27.4%) rated it 8. These high-end ratings collectively represent 73.5% of all responses, indicating that more than two-thirds of users found the platform highly useful. At the lower end, 3 users (2.7%) gave a rating of 5, while only 2 users gave ratings below 5 (one rating of 4 and one rating of 3). The very limited presence of low ratings indicates that almost all users found some meaningful value in the platform.

Figure 3 *User Evaluation of the Platform's Usefulness and Ease of Use*

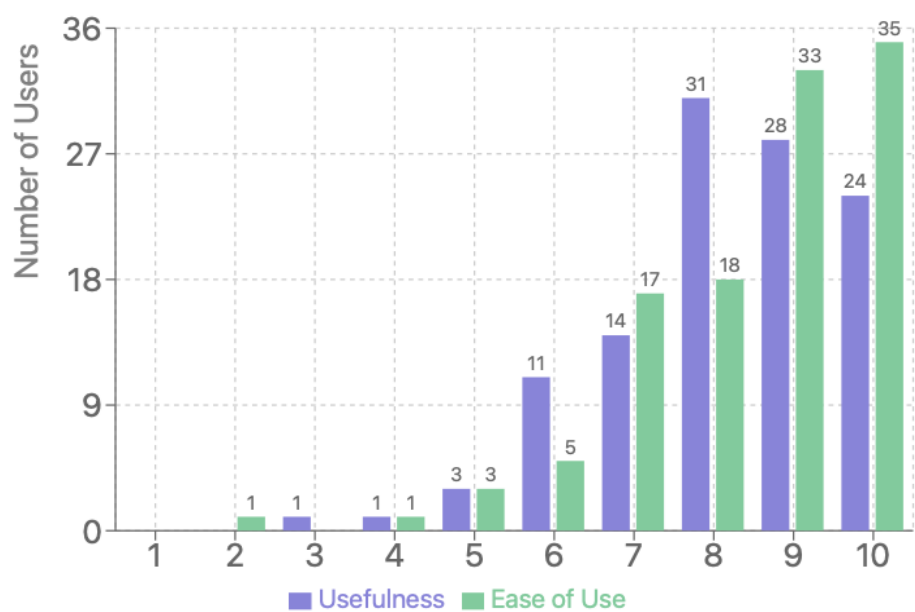

**Qualitative Analysis of User Feedback on the Platform's Usefulness**

Out of 114 participants, 98 provided written feedback on their experience with the Healthy Choice platform. A thematic analysis revealed that the participants highlighted specific features that enhanced their learning and decision-making process.

First, the platform's interactive approach to nutrition education emerged as a significant strength, with users consistently praising its ability to engage them in practical learning experiences. One participant emphasized the platform's effectiveness in real-world skill development: "I think this should become a class at our University because I personally learned so much about energy drinks and how they affect an individual's body. I am an athlete so if I had known some of these stats and facts before, it would have been amazing to just have an understanding about how drinks and food affect your body, mind, and health" (ID039). Another user expressed: "I thought the problem-solving aspect was very fun, and I liked how I did feel like a professional when looking at the nutrition facts and providing advice" (ID052).

Moreover, the simulation's authenticity was particularly valued, as noted by one user: "The scenarios are definitely relatable, and it makes sense to include relevant target audiences who use these types of beverages daily" (ID056). This realism enhanced the learning experience and made the content more relevant to users' daily lives. Another user highlighted: "I do think the scenarios are hyper-specific, so having a scenario that outlines maybe what you would choose for yourself would help you gain more insight into the priorities of students and their ability to make health-conscious decisions" (ID104).

The integration of AI assistance proved to be a particularly valuable feature that enhanced the learning experience through multiple aspects of functionality. Users frequently mentioned the real-time information seeking and clarification capabilities of AI support, with one participant noting: "I really enjoyed the ChatGPT aspect of the platform because it allowed me to ask questions I didn't know the answer to and help analyze the food labels. Sometimes I would look at the food labels and numbers and not really understand, so it was helpful being able to ask clarifying questions" (ID009). Another user stated: "I liked how AI was readily available for me to ask a basic question instead of having to open a new tab in the browser to find what I was looking for" (ID054). The AI tool's role in decision support was also highlighted: "The AI tool was also pretty useful when I wanted to ask questions I didn't know about. The real-life images and advertisements also helped to judge whether the information marketed was authentic or not" (ID074).

Lastly, the platform's comparison functionality received extensive positive feedback for its utility in decision-making. Users particularly valued the structured analysis capabilities, with one participant noting: "I found the compare feature very helpful to directly compare relevant numbers between items. I also liked that you could highlight important parts of the scenario" (ID065). The comparison tools proved especially useful for decision facilitation, as evidenced by user feedback: "The compare tool was one of the most useful tools I found when making my decision at the end" (ID021). Another user emphasized the real-world applicability of these features: "Often, at grocery stores, I get overwhelmed trying to compare product ingredients simply by looking at the labels. Having it clearly typed out, and arranged in a list form made it so much easier to compare and make an informed decision" (ID096).

**Ease of Use**

The ease of use ratings from 113 participants demonstrate a positive skew, with metrics indicating strong user satisfaction: a mean of 8.50, median of 9, and mode of 10. The distribution analysis reveals that 35 users (31.0%) gave the platform the highest possible rating of 10 for ease of use, representing the most frequent rating. This was closely followed by 33 users (29.2%) who rated it 9, and 18 users (15.9%) who gave a rating of 8. Together, these high-end ratings account for 76.1% of all responses, indicating that the majority of users found the platform very easy to use. At the lower end of the scale, similar to the usefulness ratings, 3 users (2.7%) rated it 5, 1 user (0.9%) gave it a 4, and 1 user (0.9%) rated it 2. The relatively small number of ratings below 6 (5 users, 4.4%) suggests that very few users encountered significant usability challenges.

**Qualitative Analysis of User Feedback on the Platform's Ease of Use**

Analysis of user feedback regarding the platform's ease of use reveals several key themes. The platform's basic interface structure received generally positive feedback for its simplicity. As one user noted, "The layout and user interface is simple and easy to navigate, which is a definite plus of the platform because users will feel that it is easier to learn what you're trying to convey when the platform itself is simplistic in its layout and intuitive" (ID021). Another user simply stated it was a "very easy platform to use" (2024ID005). Even users who encountered initial uncertainty quickly adapted to the interface, as one participant explained: "I thought it was easy to use after listening to the instructions and playing around with it a bit!" (ID072).

The most prominent usability challenge for users was the inability to view their collected information across different sections simultaneously. Many users expressed frustration with having to switch between tabs frequently. As one user explained, "It was difficult during the choosing portion because I had to go back to the main scenario to retrieve the necessary information. This was the same when writing the final explanation" (ID011). Another user suggested, "I would have preferred if I had the highlighted list of quotes available on the side at all times" (ID016). The qualitative feedback regarding usability challenges provides valuable insights for future platform refinement.

**Discussion**

This study evaluated a theory-driven and AI-enhanced simulation platform designed to develop nutrition literacy through interactive, scenario-based learning experiences. The findings reveal high user satisfaction with both the platform's usefulness and ease of use, suggesting that technology-enhanced simulation can effectively engage learners in nutrition education. The qualitative analysis identified four key strengths: interactive learning approach, scenario authenticity, AI assistance value, and comparison tool utility.

The strong positive user response to interactive learning experiences aligns with previous research demonstrating the superiority of active learning approaches over passive information delivery in health education. While traditional nutrition education often relies on didactic instruction, our findings support Mackert et al.'s argument that interactive, technology-enhanced approaches can better develop practical decision-making skills.[8] The positive user response to scenario-based learning extends Wickham and Carbone's work on technology in food literacy programs by demonstrating how authentic contexts can make nutrition education more relevant and engaging for learners.[10]

The integration of AI assistance represents a novel contribution to nutrition education technology. Previous studies have explored various digital tools for nutrition education, but few have investigated the role of AI in supporting learning processes. Our findings suggest that AI can serve as an effective scaffold for nutrition decision-making, providing just-in-time support that helps learners bridge knowledge gaps and develop deeper understanding. This aligns with Chiu et al.'s findings about the value of technology-embedded support in health education, while extending their work by demonstrating how AI can provide more sophisticated and adaptive assistance.[11]

The platform's effectiveness can be understood through the lens of self-regulated learning theory. The highlighting and note-taking features supported the forethought phase by helping learners identify key requirements and plan their approach. The AI assistance and product comparison tools facilitated the performance phase by enabling systematic information gathering and strategy adjustment. The final justification step prompted self-reflection, completing the self-regulated learning cycle. These findings extend Zimmerman's work by demonstrating how technology can support each phase of self-regulated learning in the context of nutrition education.[15]

The comparison functionality emerged as a particularly valuable feature, addressing a common challenge in nutrition literacy identified by Rothman et al.: the difficulty of comparing and evaluating nutritional information across products.[4] User feedback suggests that structured comparison tools can make this process more manageable and systematic, potentially leading to better-informed decisions. This finding has important implications for the design of nutrition education tools, suggesting that explicit support for product comparison should be a key feature.

Several limitations should be considered when interpreting these findings. First, the participant sample consisted of university students from a single institution, potentially limiting generalizability to other populations, particularly those with lower baseline health literacy or different demographic characteristics. Second, while the platform received positive user feedback, this study did not assess actual improvements in nutrition knowledge or decision-making abilities through pre/post measures. Third, technical challenges with some platform features, particularly AI functionality, may have affected user experiences and should be addressed in future iterations. Future research should address these limitations through longitudinal studies examining sustained learning outcomes and behavior change. Studies with diverse populations would help establish the platform's effectiveness across different demographic groups and literacy levels.

**Implications for Research, Practice and Policy**

The Healthy Choice platform's development and evaluation reveal several significant implications for research, practice, and policy in health education and nutrition literacy. In the research domain, the platform demonstrates the potential of combining scenario-based learning with artificial intelligence to create more engaging and effective health education interventions. The consistently high user ratings and qualitative feedback suggest that integrating AI assistance into educational platforms can successfully scaffold complex decision-making processes in health contexts. This opens new avenues for investigating how AI-supported learning environments can be optimized to enhance health literacy outcomes across different populations and contexts. Moreover, the design of the platform, grounded in self-regulated learning theory,[15,16] encourages learners to take an active role in their learning process. By setting goals, gathering information, monitoring their progress, and reflecting on their decisions, learners develop metacognitive skills and self-regulation strategies that are transferable to other domains of learning and life.

From a practical perspective, the platform engages users through authentic scenarios and interactive features, which provides valuable insights for health educators and instructional designers. The strong positive feedback regarding the comparison tools and AI assistance suggests that similar features should be incorporated into future health education technologies. Healthcare providers and nutrition educators could adapt this approach to create tailored interventions for specific populations or health conditions. Additionally, the platform follows modular design principles, which allow for continuous expansion and customization, making it adaptable to various educational settings and cultural contexts.

The findings also have important implications for educational policy. The user response, exemplified by requests to implement similar programs as university courses, suggests a need to integrate technology-enhanced nutrition education more broadly into educational curricula. This aligns with growing recognition of the importance of health literacy in preparing students for lifelong well-being. Educational institutions should consider incorporating similar interactive platforms into their health education programs, moving beyond traditional passive information delivery to more engaging, skill-based approaches. Moreover, the platform has implications for public health policy. Policymakers should consider supporting the development and implementation of similar tools as part of broader public health initiatives aimed at improving population-level nutrition literacy.

HEALTHY CHOICE                                                                                                                199. Coughlin SS, Whitehead M, Sheats JQ, Mastromonico J, Hardy D, Smith SA. Smartphone applications for promoting healthy diet and nutrition: a literature review. *Jacobs J Food Nutr*. 2015;2(3):021.

10. Wickham CA, Carbone ET. What's technology cooking up? A systematic review of the use of technology in adolescent food literacy programs. *Appetite*. 2018;125:333-344.

11. Chiu CJ, Kuo SE, Lin DC. Technology-embedded health education on nutrition for middle-aged and older adults living in the community. *Glob Health Promot*. 2019;26(3):80-87.

12. O'Brien BC, Battista A. Situated learning theory in health professions education research: a scoping review. *Advances in Health Sciences Education*. 2020;25:483-509.

13. Coughlan EK, Williams AM, McRobert AP, Ford PR. How experts practice: A novel test of deliberate practice theory. *J Exp Psychol Learn Mem Cogn*. 2014;40(2):449.

14. Azevedo R. Computer environments as metacognitive tools for enhancing learning. *Educ Psychol*. 2005;40(4):193-197.

15. Zimmerman BJ. Self-regulated learning and academic achievement: An overview. *Educ Psychol*. 1990;25(1):3-17.

16. Chitra E, Hidayah N, Chandratilake M, Nadarajah VD. Self-regulated learning practice of undergraduate students in health professions programs. *Front Med (Lausanne)*. 2022;9:803069.